\documentclass[12pt, prd, showpacs]{revtex4}
\usepackage{amsfonts}
\usepackage{amsmath}

\input{tcilatex}

\begin{document}

\title{Quasi-exactly solvable models based on special functions}
\author{S. N. Dolya}
\email{sdolya@gmail.com}

\begin{abstract}
We suggest a systematic method of extension of quasi-exactly solvable (QES)
systems. We construct finite-dimensional subspaces on the basis of special
functions (hypergeometric, Airy, Bessel ones) invariant with respect to the
action of differential operators of the second order with polynomial
coefficients. As a example of physical applications, we show that the known
two-photon Rabi Hamiltonian becomes quasi-exactly solvable at certain values
of parameters when it can be expressed in terms of corresponding QES
operators related to the hypergeometric function.
\end{abstract}

\maketitle

\address{{B. Verkin Institute for Low Temperature Physics and }Engineering \\
47, Lenin Prospekt, Kharkov 61164, Ukraine.}

\section{Introduction}

Usually, the models of the quantum mechanics are considered to be
exactly-solvable (when the eigenvalues and eigenfunctions are known) or
non-solvabe at all (when the eigenvalues and eidenfunctions are unknown, and
they can be found numerically only). It became a great surprise that an
intermediate case is also possible, when inside the Hilbert space there is
an invariant subspace for which the eigenvalues and eigenfunctions could be
found from algebraic equations. This type of systems is called quasi-exactly
solvable (QES). For the wide range of the one-dimensional QES the
Hamiltonian possesses hidden Lie algebra $sl(2)$ and represents a bilinear
form of first-order differential operators which satisfy the same commutator
relations as the spin ones \cite{1r}, \cite{2r}, \cite{3r}, \cite{4r}. The
QES Schr\"{o}dinger operators based on the $sl(2)$ representation were
studied in \cite{1r}, \cite{5r}, \cite{6r}, \cite{7r}, \cite{8r}. Later it
became possible to go beyond the Lie algebraic context. In particular,
starting from the second-order differential operators 
\begin{equation}
T=q(x)\cdot\frac{d^{2}}{dx^{2}}+p(x)\cdot\frac{d}{dx}+r(x)  \label{1q}
\end{equation}
such that they possess the finite-dimensional representations both in the
subspace $\mathcal{P}_{n}=span\{1,$ $x,...,$ $x^{n-1},$ $x^{n}\}$ of all
monomials of the degree $\leq n$ and in the monomial subspace span$\{1,$ $%
x^{2},...$ $,$ $x^{n-1},$ $x^{n}\}$ \cite{9r}, \cite{10r}, \cite{11r}. In 
\cite{12r}, \cite{13r},.the extension of the the corresponding subspace was
suggested 
\begin{equation}
\nu=\mathcal{P}_{n}\cup g(x)\mathcal{P}_{n}=span%
\{1,x^{2},...,x^{n-1},x^{n},g(x),g(x)\cdot x,...,g(x)\cdot x^{n}\}
\label{2q}
\end{equation}
for certain functions $g(x)$ \cite{12r}, \cite{13r}.

In the present paper we consider the problem of constructing QES operators
which preserve subspaces of a more general form:%
\begin{equation}
\mathcal{M}_{n}=span\{f_{1}(x),f_{2}(x),...,f_{n-1}(x),f_{n}(x)\}\text{.}
\label{3q}
\end{equation}
The general features of differential operators with invariant subspace $%
\mathcal{M}_{n}$ (\ref{3q}) were discussed in \cite{15r}. The main result
obtained is the following: any linear operator $P_{n}$ $\in\mathcal{P}(%
\mathcal{M}_{n})$, that preserve $\mathcal{M}_{n}$ (\ref{3q}), may be
represented as follows (a formula 5.8 in \cite{15r}): 
\begin{equation}
P_{n}=\sum_{i,j}a_{ij}\cdot f_{i}\left( x\right) \cdot L_{j}+\sum R_{m}\cdot
K_{m},  \label{4q}
\end{equation}
where $a_{ij}$ are arbitrary constants, $R_{\nu}\in\mathcal{D}$ are
arbitrary linear differential operators, $\mathcal{D}=\mathcal{D}\left( 
\mathcal{F}\right) $ is a space of the differential operators with
coefficients belonging to the functional space $\mathcal{F}$ (hereafter
denoted $\mathcal{P}_{n}$), $K_{m}$ are operators belonging to the
annihilator $\mathcal{A}\left( \mathcal{M}_{n}\right) $: 
\begin{equation}
\mathcal{A}\left( \mathcal{M}_{n}\right) =\{K\in\mathcal{D}\text{ }|\text{ }%
K\left( f(x)\right) =0\text{, for all }f(x)\in\mathcal{M}_{n}\},  \label{5q}
\end{equation}
$L_{j}-$ are operators belonging to the affine annihilator $\mathcal{K}%
\left( M_{n}\right) $: 
\begin{equation}
\mathcal{K}\left( \mathcal{M}_{n}\right) =\{L\in\mathcal{D}\text{ }|\text{ }%
L\left( f(x)\right) =c\in\mathbb{R}\text{, for all }f(x)\in\mathcal{M}_{n}\}
\label{6q}
\end{equation}
$c$ is a constant. All the definitions and symbols are in agreement with the
ones used in \cite{15r}.

In spite of the fact that the theorem of existence specifies a general form
of the required differential operators (\ref{4q}), their finding is,
generally speaking, not trivial task. The purpose of the given work consists
in explicit construction of such QES operators, for which the set of
functions $f_{n}(x)$ does not reduce to polynomials and represents special
functions. With this purpose, we offer a method of QES extension. It
consists in the following. To construct the QES Hamiltonians on the basis of
the subspace (\ref{3q}), we select in (\ref{6q}) the subspace of
differential operators $\mathcal{K}_{2}\left( \mathcal{M}_{n}\right) $ of
the degree two or less: 
\begin{equation}
\mathcal{K}_{2}\left( \mathcal{M}_{n}\right) =\{L\in\mathcal{K}\left( 
\mathcal{M}_{n}\right) \text{ }|\text{ }order(L)\leq2\}\text{.}  \label{7q}
\end{equation}
The general approach suggested in \cite{15r} does not supply us with
information, whether or not the space $\mathcal{K}_{2}\left( \mathcal{M}%
_{n}\right) $ (\ref{7q}) is empty, without the constructing $\mathcal{K}%
\left( \mathcal{M}_{n}\right) $ for the chosen $\mathcal{M}_{n}$ (\ref{3q}).
To simplify matter, we start from the two-dimensional subspace $\mathcal{M}%
_{2}=span\{f_{1}(x),f_{2}(x)\}$, for which operators $L_{i}\in\mathcal{K}%
_{2}\left( \mathcal{M}_{2}\right) $ are known to exist and can be explicitly
constructed \cite{15r}. Let us try to extend the subspace $\mathcal{M}_{2}$
by adding function $f_{3}(x)$ to it in such a way that the new subspace $%
\mathcal{M}_{3}=span\{f_{1}(x),f_{2}(x),f_{3}(x)\}$ be the three-dimensional
and the space of corresponding operators $\mathcal{K}_{2}\left( \mathcal{M}%
_{3}\right) $ (\ref{7q}) be non-empty. Thus, transformation $\mathcal{M}%
_{2}\longrightarrow\mathcal{M}_{3}$ has to lead to the change of the
coefficients at $\frac{d}{dx}$ in the operators $L_{i}\in\mathcal{K}%
_{2}\left( \mathcal{M}_{2}\right) $, with the order of the operators $%
L_{i}\in\mathcal{K}_{2}\left( \mathcal{M}_{2}\right) $ preserved: $%
order(L_{i})\leq2$. Repeating this operation $n$ times, we obtain the
QES-extension of $\mathcal{M}_{2}$. In other words, the initial subspace $%
\mathcal{M}_{2}=span\{f_{1}(x),f_{2}(x)\}$ permits the QES-extension if $%
\mathcal{M}_{2}$ may be extended to $\mathcal{M}_{n}$ under the condition
that the affine annihilator $\mathcal{K}_{2}\left( \mathcal{M}_{n}\right) $ (%
\ref{7q}) is not empty. (It is worth noting another method of constructing
invariant subspaces (\ref{3q}) which is based on the conditional symmetries 
\cite{14r}, but it will not be discussed in the present paper.)

\section{Construction of the invariant subspaces}

Below we suggest the method that enables one to construct invariant
subspaces $\mathcal{M}_{n}$. Let us start from the simplest case $n=2$. Let
us choose a linear independent basis $\{f_{0}^{+}(x),f_{0}^{-}(x)\}$ for the
subspace $\mathcal{M}_{2}=span\{f_{0}^{+}(x),$ $f_{0}^{-}(x)\}$ in the
following way. Let us consider the function $f(x)$, which satisfies the
second order homogeneous differential equation with polynomial coefficients: 
$q(x)\frac {d^{2}}{dx^{2}}f(x)+p(x)\frac{d}{dx}f(x)+r(x)f(x)=0$. The case
when $f(x)$ $=$ $constant\cdot\frac{d}{dx}f(x)$ is eliminated. We select the
function $f_{0}^{+}(x)=f(x)$ and its derivative $f_{0}^{-}(x)=\frac{d}{dx}%
f(x)=f^{\prime}(x)$ as the basis components of $\mathcal{M}_{2}$: 
\begin{equation}
\mathcal{M}_{2}=span\{f_{0}^{+}(x),f_{0}^{-}(x)\}\text{.}  \label{8q}
\end{equation}
For simplicity, we shall work directly with the operator (\ref{4q}) instead
of the operators $L_{i}$ representing affine annihilator $\mathcal{K}\left( 
\mathcal{M}_{2}\right) $ (\ref{6q}).

\bigskip Our strategy can be described as follows:

\begin{itemize}
\item[I] We find a general form of the operator of the second order $P_{2}$
for which subspace $\mathcal{M}_{2}$ (\ref{9q}) is preserved. The existence
of such an operator is supported of Lemma 4.10 \cite{15r}.

\item[II] We make extension of the subspace $\mathcal{M}_{2}\rightarrow 
\mathcal{M}_{4}=span\{f_{0}^{+},f_{1}^{+},f_{0}^{-},f_{1}^{-}\}$.

\item[III] We find a general form of the operator of the second order $P_{4}$
for which subspace $\mathcal{M}_{4}$ is preserved. If a non-trivial solution 
$P_{4}\neq const$ exists for the chosen way of the extension, we pass to
item IV, otherwise we change a way of extension.

\item[IV] We make comparison of two operators $P_{2}$ and $P_{4}$ and try to
guess a general form of coefficients which enable us to repeat extension in
the chain ( $\mathcal{M}_{2}$ $\rightarrow$ $\mathcal{M}_{4}\rightarrow$ $%
... $ $\rightarrow$ $\mathcal{M}_{2\left( N+1\right) }$) ($N=0,1,2,...$) )
that leaves these subspaces invariants. As a result, we obtain the explicit
form of operator $P_{2\left( N+1\right) }$ that acts on the elements of the
subspace $\mathcal{M}_{2\left( N+1\right) }=\{f_{0}^{+},$ $f_{1}^{+},$ $%
f_{2}^{+}...$ $f_{N}^{+},$ $f_{0}^{-},$ $f_{1}^{-},$ $f_{2}^{-},...,$ $%
f_{N}^{-}\}$, $N=0,1,2,...$. This is not the end of story since the general
form does not fix by itself the concrete expression for the coefficients.

\item[V] With the guessed general form of the operator $P_{2\left(
N+1\right) }$ at hand, we require that it leave the corresponding subspace
invariant, whence we find the concrete values of its coefficients.
\end{itemize}

For explicit demonstration of the aforementioned algorithm, we shall
consider a concrete choice of the function $f_{0}^{+}(x):f_{0}^{+}(x)=%
\begin{array}{c}
_{0}F_{1}\left[ \QATOP{-}{s};x\right]%
\end{array}
$ \cite{16r} where $%
\begin{array}{c}
_{0}F_{1}\left[ \QATOP{-}{s};x\right]%
\end{array}
$ is a hypergeometric function, $f_{0}^{-}(x)=%
\begin{array}{c}
_{0}F_{1}\left[ \QATOP{-}{s+1};x\right]%
\end{array}
$ is its derivative up to a constant factor.

The differential equation which the hypergeometric function obeys $\left(
x\cdot\frac{d^{2}}{dx^{2}}+s\cdot\frac{d}{dx}-1\right) f_{0}^{+}(x)=0$ as
well as its other properties can be found, e.g., in \cite{16r}.Then, we
have: 
\begin{equation}
\mathcal{M}_{2}=span\left\{ 
\begin{array}{c}
_{0}F_{1}\left[ \QATOP{-}{s};x\right]%
\end{array}
,%
\begin{array}{c}
_{0}F_{1}\left[ \QATOP{-}{s+1};x\right]%
\end{array}
\right\} ,  \label{9q}
\end{equation}

Let us choose the operator of the second order $P_{2}=a_{3}(x)\frac{d^{2}}{%
dx^{2}}+a_{2}(x)\frac{d}{dx}+a_{0}(x)$ and write down the condition of
invariance of the subspace $\mathcal{M}_{2}$ (\ref{9q}) for the given
operator: 
\begin{equation}
P_{2}\left( f_{0}^{+}\right) =c_{1}\cdot f_{0}^{+}+c_{2}\cdot f_{0}^{-}
\label{10nq}
\end{equation}%
\begin{equation}
P_{2}\left( f_{0}^{-}\right) =c_{3}\cdot f_{0}^{+}+c_{4}\cdot f_{0}^{-}
\label{11nq}
\end{equation}
Using rules of differentiation $\frac{d}{dx}$ $f_{0}^{+}=\frac{f_{0}^{+}}{x}-%
\frac{f_{0}^{-}}{x}$, $\frac{d}{dx}$ $f_{0}^{-}=-s\left( s+1\right) \frac{%
f_{0}^{+}}{x^{2}}$ $+$ $\left( s+s^{2}+x\right) \frac{f_{0}^{-}}{x^{2}}$ and
equating coefficients at functions $f_{0}^{+}$, $f_{0}^{-}$ (\ref{10nq}, \ref%
{11nq}) we obtain the solution:%
\begin{align}
a_{3}(x) & =-c_{3}\cdot\frac{x^{2}}{s}+c_{2}\cdot s\cdot x,  \notag \\
a_{2}(x) & =-c_{3}\cdot x+c_{2}\cdot s\left( s+1\right) ,  \label{12nq2} \\
a_{1}(x) & =c_{3}\cdot\frac{x}{s}-c_{2}\cdot s+c_{4}+c_{3}  \notag
\end{align}
$c_{1}=c_{4}+c_{3}$. Substituting obtained coefficients $a_{k}(x)$, ($%
k=1,2,3 $) (\ref{12nq2}) into the expression for the operator $P_{2}$ we
have:%
\begin{equation}
P_{2}=c_{2}\cdot s\cdot P_{2}^{1}-\frac{c_{3}}{s}\cdot
P_{2}^{2}+c_{4}+c_{3}+s\cdot c_{2}  \label{13qn}
\end{equation}
where%
\begin{equation}
P_{2}^{1}=x\frac{d^{2}}{dx^{2}}+\left( s+1\right) \cdot\frac{d}{dx}\text{, }%
P_{2}^{2}=x^{2}\frac{d^{2}}{dx^{2}}+s\cdot x\frac{d}{dx}-x.  \label{14qn}
\end{equation}
One can check that operators $P_{2}^{1}$, $P_{2}^{2}$ act on the basis
functions according to the formulas $P_{2}^{1}\left( f_{0}^{+}\right)
=f_{0}^{+}+\frac{1}{s}\cdot f_{0}^{-}$, $P_{2}^{1}\left( f_{0}^{-}\right)
=f_{0}^{-}$, $P_{2}^{2}\left( f_{0}^{+}\right) =0$, $P_{2}^{2}\left(
f_{0}^{-}\right) =s\cdot f_{0}^{-}-s\cdot f_{0}^{+}$ and, thus, preserve the
subspace $\mathcal{M}_{2}$ (\ref{9q}). This is agreement with the fact that
the affine annihilator $\mathcal{K}_{2}\left( \mathcal{M}\right) $ has the
property of the vector space and possesses the operator basis (\ref{14qn})
(see in \cite{15r}).

The extension of the subspace $\mathcal{M}_{2}$ (\ref{9q}) can be realized
in many ways. The basic requirement here consists in that after adding new
elements of the basis $\mathcal{M}_{2}$ $\cup$ $span\left\{
f_{1}^{+},f_{1}^{-}\right\} $ $=$ $span\left\{
f_{0}^{+},f_{1}^{+},f_{0}^{-},f_{1}^{-}\right\} $ $=$ $\mathcal{M}_{4}$ the
operator basis of $P_{2}$ (\ref{13qn}, \ref{14qn}) should remain
two-dimensional for the new subspace $\mathcal{M}_{4}$. We consider the
simplest way of extension - multiplication of the basis functions by the
power function $x^{n}$: $\mathcal{M}_{2}\rightarrow\mathcal{M}%
_{4}\rightarrow...\rightarrow\mathcal{M}_{2\left( N+1\right) }$ ($%
N=0,1,2,....$), 
\begin{equation}
\{f_{0}^{+},f_{0}^{-}\}\rightarrow\{f_{0}^{+},f_{1}^{+},f_{0}^{-},f_{1}^{-}%
\}\rightarrow...\rightarrow%
\{f_{0}^{+},f_{1}^{+},...f_{N}^{+},f_{0}^{-},f_{1}^{-},...,f_{N}^{-}\}
\label{16qn}
\end{equation}
where 
\begin{equation}
f_{n}^{+}=x\cdot f_{n-1}^{+}=x^{n}\cdot f_{0}^{+},f_{n}^{-}=x\cdot
f_{n-1}^{-}=x^{n}\cdot f_{0}^{-},  \label{xn}
\end{equation}
$n=0,$ $1,..,$ $N-1,$ $N$. To demonstrate that suggested extension is
suitable, we shall consider subspace $\mathcal{M}_{4}=\left\{
f_{0}^{+},f_{1}^{+},f_{0}^{-},f_{1}^{-}\right\} $ and shall write down the
condition of its invariance with respect to the operator $P_{4}=b_{3}(x)%
\frac{d^{2}}{dx^{2}}+b_{2}(x)\frac{d}{dx}+b_{0}(x)$:%
\begin{equation}
P_{4}\left( \overrightarrow{f}\right) =\widehat{d}\cdot\overrightarrow{f}
\label{17qn}
\end{equation}
where $\overrightarrow{f}=\left\{
f_{0}^{+},f_{1}^{+},f_{0}^{-},f_{1}^{-}\right\} ^{T}$ - is vector, $\widehat{%
d}=[d_{i,j}]$, ($i=1,...,4$; $j=1,...,4$) - is the matrix of coefficients.
The condition (\ref{17qn}) represents the system of the linear equations on
functions $b_{k}(x)$, $k=1,$ $2,$ $3$ and quantities $d_{i,j}$, ($i=1,...,4$%
; $j=1,...,4$). Solving this linear system of the equations we have:%
\begin{align}
b_{3}(x) & =d_{13}\cdot s\cdot x-d_{14}\cdot s\frac{x^{2}}{2},  \label{18qn}
\\
b_{2}(x) & =d_{13}\cdot s\left( s+1\right) +d_{14}\cdot x\cdot s\left( 1-%
\frac{s}{2}\right) ,  \label{19qn} \\
b_{1}(x) & =d_{13}\cdot\left( -s\right) +d_{14}\cdot\left( \frac{sx}{2}%
+s^{2}-s+1\right)  \label{20qn}
\end{align}%
\begin{equation}
d_{11}=d_{14}\cdot\left( s^{2}-s\right) +d_{44},d_{21}=d_{13}\cdot\left(
s^{2}+s\right) ,d_{22}=d_{14}\cdot\frac{s^{2}}{2}+d_{44},d_{24}=3d_{13},
\end{equation}%
\begin{equation}
d_{31}=d_{14}\cdot\frac{3s^{2}}{2},d_{33}=-d_{14}\cdot\left( \frac{s^{2}}{2}%
+s\right) +d_{44},d_{41}=2d_{13}\cdot s^{2},
\end{equation}%
\begin{equation}
d_{42}=d_{14}\cdot\frac{s^{2}}{2},d_{43}=d_{13}\cdot\left( s-s^{2}\right)
,d_{12}=d_{23}=d_{32}=d_{34}=0.
\end{equation}
The existence of the non-trivial solution (\ref{18qn}-\ref{20qn}) means that
the way by which extension (\ref{16qn}) was made is suitable.

Let us analyze the general form of the operator $P_{4}$ for the obtained
solution:%
\begin{equation}
P_{4}=d_{13}\cdot s\cdot P_{4}^{1}-\frac{d_{14}\cdot s}{2}%
P_{4}^{2}+d_{44}-d_{13}\cdot s+d_{14}\cdot\left( s^{2}-s\right)  \label{21qn}
\end{equation}
where%
\begin{equation}
P_{4}^{1}=x\frac{d^{2}}{dx^{2}}+\left( s+1\right) \cdot\frac{d}{dx}\text{, }%
P_{4}^{2}=x^{2}\frac{d^{2}}{dx^{2}}+\left( s-2\right) \cdot x\frac{d}{dx}-x.
\label{22qn}
\end{equation}
The operator $P_{4}$ (\ref{21qn}) as well as operator $P_{2}$ (\ref{13qn}),
up to an additive constant , depends on two free parameters $d_{13}$, $%
d_{14} $ (\ref{21qn}). From the general solution (\ref{21qn}) we select two
operators $P_{4}^{1}$, $P_{4}^{2}$ (\ref{22qn}) which do not depend on
parameters $d_{13}$, $d_{14}$, $d_{44}$. We \ see that the coefficients of
the operator $P_{4}^{1}$ coincide with those of the operator $P_{2}^{1}$ (%
\ref{14qn}). This suggests the idea to implement the expression (\ref{22qn})
for the operator $P_{2\left( N+1\right) }^{1}$ for any $N$ and subspace (\ref%
{16qn}). The situation with the operator $P_{m}^{2}$ is somewhat more
complicated. The natural general guessed form for this operator reads%
\begin{equation}
P_{2\left( N+1\right) }^{2}=x^{2}\frac{d^{2}}{dx^{2}}+\left( s-\alpha
_{N}\right) \cdot x\frac{d}{dx}-x  \label{23qn}
\end{equation}
where the quantity $\alpha_{N}$ depends on $N$ and, besides, satisfies the
conditions $\alpha_{0}=0$, $\alpha_{1}=2$ as it follows from (\ref{14qn}, %
\ref{22qn}). To find dependence $\alpha_{N}$ from $N$, we shall consider the
result of the action of the operator $P_{2\left( N+1\right) }^{2}$ on the
element $f_{n}^{+}$ (\ref{16qn}):%
\begin{equation}
P_{2\left( N+1\right) }^{2}\left( f_{n}^{+}\right) =n\left(
n+s-1-\alpha_{N}\right) \cdot f_{n}^{+}+\frac{\left( 2n-\alpha_{N}\right) }{s%
}\cdot f_{n+1}^{-}.  \label{24qn}
\end{equation}
If we put $n=N$ in (\ref{24qn}) and require that the subspace be
finite-dimensional, the condition of cut off at $n=N$ gives us $a_{N}=2\cdot
N$, whence the operator $P_{2\left( N+1\right) }^{2}$ looks like $P_{2\left(
N+1\right) }^{2}=x^{2}\frac{d^{2}}{dx^{2}}+\left( s-2\cdot N\right) \cdot x%
\frac{d}{dx}-x$. As a result, we obtained a pair of operators that indeed
leave the subspace $\mathcal{M}_{2\left( N+1\right) }$ invariant.
Correspondingly, any linear combination $P_{2\left( N+1\right)
}=\kappa_{1}\cdot P_{2\left( N+1\right) }^{1}$ $+$ $\kappa_{2}\cdot
P_{2\left( N+1\right) }^{2}$ where $\kappa_{1}$, $\kappa_{2}$ arbitrary
constants has the same property.

Following the way we described above, we found a series of concrete examples
1, 2, 4, 5, 6. The list of obtained finite-dimensional subspaces together
with the basic operators for which corresponding subspaces are invariant is
given below. Meanwhile, the suggested way of the QES - extension (\ref{xn})
of the operators (\ref{4q}) is not unique. As the alternative approach, in
an example 3 we used another "natural" procedure of QES - extension in that
an integer $n$ labeling the basis was added to the parameter of the
function: $\alpha\rightarrow\alpha+n$.

\textbf{1}) Finite dimensional function subspace $\mathcal{R}%
_{N}^{1}=span\{f_{0}^{+},$ $...,$ $f_{N}^{+},$ $f_{0}^{-},$ $...,$ $%
f_{N}^{-}\}$ $(N=0,1,2,....),$ $\dim\left( \mathcal{R}_{N}^{1}\right)
=2\left( N+1\right) $, formed by functions $f_{n}^{+}=x^{n}\cdot%
\begin{array}{c}
_{0}F_{1}\left[ \QATOP{-}{s};x\right]%
\end{array}
$, $f_{n}^{-}=x^{n}\cdot%
\begin{array}{c}
_{0}F_{1}\left[ \QATOP{-}{s+1};x\right]%
\end{array}
$ $(n=0,1,..,N-1,N)$ \cite{16r}, is invariant for the operators $%
J_{1}^{+}\equiv P_{2(N+1)}^{1}$, $J_{1}^{-}\equiv P_{2(N+1)}^{2}$: 
\begin{equation}
\QATOP{J_{1}^{-}=x\frac{d^{2}}{dx^{2}}+\left( s+1\right) \frac {d}{dx}}{%
\text{ }J_{1}^{+}=x^{2}\frac{d^{2}}{dx^{2}}+\left( s-2N\right) \cdot x\frac{d%
}{dx}-x}  \label{11q}
\end{equation}
The operators $J_{1}^{+}$, $J_{1}^{-}$ act on the functions $f_{n}^{+}(x)$, $%
f_{n}^{-}(x)\in\mathcal{R}_{N}^{1}$ as follows: 
\begin{align}
J_{1}^{+}\binom{f_{n}^{+}}{f_{n}^{-}} & =\binom{n\left( A_{n}-1+s\right)
\cdot f_{n}^{+}+\frac{2B_{n}}{s}\cdot f_{n+1}^{-}}{\left( n-s\right) \left(
A_{n}-1\right) \cdot f_{n}^{-}+s\left( 2B_{n}-1\right) \cdot f_{n}^{+}} 
\notag \\
J_{1}^{-}\binom{f_{n}^{+}}{f_{n}^{-}} & =\binom{f_{n}^{+}+n\left( n+s\right)
\cdot f_{n-1}^{+}+\frac{1+2n}{s}\cdot f_{n}^{-}}{f_{n}^{-}+n\left(
n-s\right) \cdot f_{n-1}^{-}+2ns\cdot f_{n-1}^{+}}  \label{12q}
\end{align}
, where $A_{n}=n-2N$, $B_{n}=n-N$. The function $f_{0}^{+}\left( x\right) $
satisfies the differential equation $\left( x\frac{d^{2}}{dx^{2}}+s\frac {d}{%
dx}-1\right) f_{0}^{+}\left( x\right) =0$.

\textbf{2}) Finite dimensional function subspace $\mathcal{R}%
_{N}^{2}=span\{f_{0}^{+},$ $...,$ $f_{N}^{+},$ $f_{0}^{-},$ $...,$ $%
f_{N}^{-}\}$ $(N=0,1,2,....),$ $\dim\left( \mathcal{R}_{N}^{2}\right)
=2\left( N+1\right) $, formed by functions $f_{n}^{+}=x^{n}\cdot%
\begin{array}{c}
_{1}F_{1}\left[ \QATOP{\alpha}{s};x\right]%
\end{array}
$, $f_{n}^{-}=x^{n}\cdot%
\begin{array}{c}
_{1}F_{1}\left[ \QATOP{\alpha+1}{s+1};x\right]%
\end{array}
$ $(n=0,1,..,N-1,N)$ \cite{16r}, is invariant for the operators $%
J_{2}^{+},J_{2}^{-}$: 
\begin{equation}
\QATOP{J_{2}^{-}=x\frac{d^{2}}{dx^{2}}+\left( 1+s-x\right) \frac {d}{dx}}{%
\text{ }J_{2}^{+}=x^{2}\frac{d^{2}}{dx^{2}}+\left( s-2N-x\right) \cdot x%
\frac{d}{dx}+x\left( N-\alpha\right) }  \label{13q}
\end{equation}
The operators $J_{2}^{+},$ $J_{2}^{-}$ act on the functions $f_{n}^{+}(x),$ $%
f_{n}^{-}(x)\in\mathcal{R}_{N}^{2}$ as follows: 
\begin{align}
J_{2}^{+}\binom{f_{n}^{+}}{f_{n}^{-}} & =\binom{n\left( A_{n}-1+s\right)
\cdot f_{n}^{+}-B_{n}\cdot f_{n+1}^{+}+2\alpha B_{n}/s\cdot f_{n+1}^{-}}{%
\left( s-n\right) \left( 1-A_{n}\right) \cdot f_{n}^{-}+B_{n}\cdot
f_{n+1}^{-}+s\left( 2B_{n}-1\right) \cdot f_{n}^{+}}  \label{14qm} \\
J_{2}^{-}\binom{f_{n}^{+}}{f_{n}^{-}} & =\binom{\left( \alpha-n\right)
f_{n}^{+}+n\left( n+s\right) \cdot f_{n-1}^{+}+\alpha\left( 1+2n\right)
/s\cdot f_{n}^{-}}{\left( \alpha+n+1\right) f_{n}^{-}+n\left( n-s\right)
\cdot f_{n-1}^{-}+2ns\cdot f_{n-1}^{+}}  \label{15q}
\end{align}
, where $A_{n}=n-2N$, $B_{n}=n-N$. The function $f_{0}^{+}\left( x\right) $
satisfies the differential equation $\left( x\frac{d^{2}}{dx^{2}}+\left(
s-x\right) \frac{d}{dx}-\alpha\right) f_{0}^{+}\left( x\right) =0$.

\textbf{3}) Finite dimensional function subspace $\mathcal{R}%
_{N}^{3}=span\{f_{0},$ $f_{1},$ $...,$ $f_{N}\}$ $(N=0,1,2,....),$ $%
\dim\left( \mathcal{R}_{N}^{3}\right) =N+1$, formed by functions $f_{n}=%
\begin{array}{c}
_{1}F_{1}\left[ \QATOP{\alpha+n}{s};x\right]%
\end{array}
$ $(n=0,1,..,N-1,N)$ \cite{16r}, is invariant for the operators $%
J_{3}^{+},J_{3}^{-}$: 
\begin{equation}
\QATOP{J_{3}^{-}=x\frac{d^{2}}{dx^{2}}+\left( s-x\right) \frac {d}{dx}}{%
\text{ }J_{3}^{+}=x^{2}\frac{d^{2}}{dx^{2}}+\left( s-N-x\right) \cdot x\frac{%
d}{dx}-\alpha x}  \label{16q}
\end{equation}
The operators $J_{3}^{+},$ $J_{3}^{-}$ act on the functions $f_{n}(x)\in%
\mathcal{R}_{N}^{3}$ as follows: 
\begin{align}
J_{3}^{-}\left( f_{n}\right) & =\left( n+\alpha\right) \cdot f_{n}
\label{17q} \\
J_{3}^{+}\left( f_{n}\right) & =\left( sn+\left( \alpha+n\right)
C_{n}\right) \cdot f_{n}+\left( \alpha+n\right) B_{n}\cdot f_{n+1}+n\left(
\alpha+n-s\right) \cdot f_{n-1}  \notag
\end{align}
, where $C_{n}=N-2n$, $B_{n}=n-N$ . It is worthwhile to note that
replacement $\left( s\longrightarrow s-1\text{, }\alpha\longrightarrow\alpha-%
\frac{1}{2}+\frac{N}{2}\text{, }N\longrightarrow\frac{N}{2}-\frac{1}{2}%
\right) $ transforms the operators $J_{2}^{+},$ $J_{2}^{-}$ into $J_{3}^{+},$
$J_{3}^{-}$ , if $N$ is odd.

\textbf{4}) Finite-dimensional function subspace $\mathcal{R}%
_{N}^{4}=span\{f_{0}^{+},$ $...,$ $f_{N}^{+},$ $f_{0}^{-},$ $...,$ $%
f_{N}^{-}\}$ $(N=0,1,2,....),$ $\dim\left( \mathcal{R}_{N}^{4}\right)
=2\left( N+1\right) $, formed by functions $f_{n}^{+}=x^{n}\cdot Airy\left(
x\right) ,$ $f_{n}^{-}=x^{n}\cdot\frac{d}{dx}Airy\left( x\right) $ $%
(n=0,1,..,N-1,N)$ \cite{16r}, is invariant for the operators $%
J_{4}^{+},J_{4}^{-}$: 
\begin{equation}
\QATOP{J_{4}^{-}=x\frac{d^{2}}{dx^{2}}-\left( 1+2N\right) \frac {d}{dx}-x^{2}%
}{\text{ }J_{4}^{+}=\frac{d^{2}}{dx^{2}}-x}  \label{18q}
\end{equation}
The operators $J_{4}^{+},$ $J_{4}^{-}$ act on the functions $f_{n}^{+}(x),$ $%
f_{n}^{-}(x)\in\mathcal{R}_{N}^{4}$ as follows:

\begin{equation}
J_{4}^{+}\binom{f_{n}^{+}}{f_{n}^{-}}=\binom{n\left( n-1\right) \cdot
f_{n-2}^{+}+2n\cdot f_{n-1}^{-}}{\left( 1+2n\right) \cdot f_{n}^{+}+n\left(
n-1\right) \cdot f_{n-2}^{-}}  \label{19q}
\end{equation}

\begin{equation}
J_{4}^{-}\binom{f_{n}^{+}}{f_{n}^{-}}=\binom{\left( A_{n}-2\right) n\cdot
f_{n-1}^{+}+\left( 2B_{n}-1\right) \cdot f_{n}^{-}}{2B_{n}\cdot
f_{n+1}^{+}+\left( A_{n}-2\right) n\cdot f_{n-1}^{-}}  \label{20q}
\end{equation}
where $A_{n}=n-2N$, $B_{n}=n-N$. The function $f_{0}^{+}\left( x\right) $ to
satisfy the differential equation $\left( \frac{d^{2}}{dx^{2}}-x\right)
f_{0}^{+}\left( x\right) =0$.

\textbf{5}) Finite dimensional function subspace $\mathcal{R}%
_{N}^{5}=span\{f_{0}^{+},$ $...,$ $f_{N}^{+},$ $f_{0}^{-},$ $...,$ $%
f_{N}^{-}\}$ $(N=0,1,2,....),$ $\dim\left( \mathcal{R}_{N}^{5}\right)
=2\left( N+1\right) $, formed by functions $f_{n}^{+}=x^{n}\cdot
BesselK\left( \nu,x\right) ,$ $f_{n}^{-}=x^{n}\cdot BesselK\left(
\nu+1,x\right) $ $(n=0,1,..,N-1,N)$ \cite{16r}, is invariant for the
operators $J_{5}^{+},J_{5}^{-}$: 
\begin{equation}
\QATOP{J_{5}^{-}=x\frac{d^{2}}{dx^{2}}+2\frac{d}{dx}-\frac{\nu ^{2}+\nu+x^{2}%
}{x}}{\text{ }J_{5}^{+}=x^{2}\frac{d^{2}}{dx^{2}}+x\left( 1-2N\right) \frac{d%
}{dx}-x^{2}}  \label{21q}
\end{equation}
The operators $J_{5}^{+},$ $J_{5}^{-}$ act on the functions $f_{n}^{+}(x),$ $%
f_{n}^{-}(x)\in\mathcal{R}_{N}^{5}$ as follows: 
\begin{equation}
J_{5}^{+}\binom{f_{n}^{+}}{f_{n}^{-}}=\binom{\left( \nu+n\right) \left(
\nu+A_{n}\right) \cdot f_{n}^{+}-2B_{n}\cdot f_{n+1}^{-}}{\left(
n-1-\nu\right) \left( A_{n}-1-\nu\right) \cdot f_{n}^{-}-2B_{n}\cdot
f_{n+1}^{+}}  \label{22q}
\end{equation}%
\begin{equation}
J_{5}^{-}\binom{f_{n}^{+}}{f_{n}^{-}}=\binom{n\left( 1+n+2\nu\right) \cdot
f_{n-1}^{+}-\left( 1+2n\right) \cdot f_{n}^{-}}{-\left( 1+2n\right) \cdot
f_{n}^{+}+n\left( n-2\nu-1\right) \cdot f_{n-1}^{-}}  \label{23q}
\end{equation}
, where $A_{n}=n-2N,B_{n}=n-N$. The function $f_{0}^{+}\left( x\right) $
satisfies the differential equation $\left( x^{2}\frac{d^{2}}{dx^{2}}+x\frac{%
d}{dx}-\left( x^{2}+\nu^{2}\right) \right) f_{0}^{+}\left( x\right) =0$.

\textbf{6}) Finite dimensional function subspace $\mathcal{R}%
_{N}^{6}=span\{f_{0}^{+},$ $...,$ $f_{N}^{+},$ $f_{0}^{-},$ $...,$ $%
f_{N}^{-}\}$ $(N=0,1,2,....),$ $\dim\left( \mathcal{R}_{N}^{6}\right)
=2\left( N+1\right) $, formed by functions $f_{n}^{+}=x^{n}\cdot%
\begin{array}{c}
_{1}F_{1}\left[ \QATOP{\alpha}{1/2};x^{2}\right]%
\end{array}
,$ $f_{n}^{-}=x^{n}\cdot%
\begin{array}{c}
_{1}F_{1}\left[ \QATOP{\alpha+1}{3/2};x^{2}\right]%
\end{array}
$ $(n=0,1,..,N-1,N)$ \cite{16r}, is invariant for the operators $%
J_{6}^{+},J_{6}^{-}$: 
\begin{equation}
\QATOP{J_{6}^{-}=\frac{d^{2}}{dx^{2}}-2x\frac{d}{dx}}{\text{ }J_{6}^{+}=x%
\frac{d^{2}}{dx^{2}}-\left( 2x^{2}+1+2N\right) \frac{d}{dx}+2x\left(
N-2\alpha\right) }  \label{24q}
\end{equation}
The operators $J_{6}^{+},$ $J_{6}^{-}$ act on the functions $f_{n}^{+}(x)$, $%
f_{n}^{-}(x)\in\mathcal{R}_{N}^{6}$ as follows: 
\begin{equation}
J_{6}^{+}\binom{f_{n}^{+}}{f_{n}^{-}}=\binom{-2B_{n}\cdot
f_{n+1}^{+}+n\left( A_{n}-2\right) \cdot f_{n-1}^{+}+4\alpha\left(
2B_{n}-1\right) \cdot f_{n}^{-}}{n\left( A_{n}-2\right) \cdot
f_{n-1}^{-}+\left( 2B_{n}-1\right) \cdot f_{n}^{+}+2B_{n}\cdot f_{n+1}^{-}}
\label{25q}
\end{equation}

\begin{equation}
J_{6}^{-}\binom{f_{n}^{+}}{f_{n}^{-}}=\binom{2\left( 2\alpha-n\right) \cdot
f_{n}^{+}+\left( n^{2}-n\right) \cdot f_{n-2}^{+}+8\alpha n\cdot f_{n-1}^{-}%
}{2n\cdot f_{n-1}^{+}+2\left( 2\alpha+1+n\right) \cdot f_{n}^{-}+\left(
n^{2}-n\right) \cdot f_{n-2}^{-}}  \label{26q}
\end{equation}
where $A_{n}=n-2N$, $B_{n}=n-N$.

\qquad For the proof of equalities (\ref{12q}, \ref{14qm}, \ref{15q}, \ref%
{17q}, \ref{19q}, \ref{20q}, \ref{22q}, \ref{23q}, \ref{25q}, \ref{26q}) it
is enough to use rules of differentiation of special functions or their
representations by the power series \cite{16r}. The commutation rules for
the operators $J_{k}^{+},$ $J_{k}^{-}$ $(k=1\ldots6)$ are given in appendix
A. Let us now consider an explicit example of application of the constructed
subspaces to a physical system - two-photon Rabi Hamiltonian.

\section{QES two-photon Rabi Hamiltonian}

The two-photon Rabi Hamiltonian (TPRH) is an obvious extension of the
original Rabi Hamiltonian \cite{18r}, which takes into account the atomic
transitions induced by the absorption and emission of two photons rather
than one \cite{17r}. The corresponding system as a whole is not integrable.
Let us prove that (\ref{16q}, \ref{17q}), at some choice of parameters, is
invariant subspace for TPRH: 
\begin{equation}
H=\frac{\omega_{0}}{2}\sigma_{z}+\omega\cdot b^{+}b+g\left( b^{2}+\left(
b^{+}\right) ^{2}\right) \cdot\left( \sigma_{+}+\sigma_{-}\right)
\label{27q}
\end{equation}
where $\sigma_{z},\sigma_{\pm}=\sigma_{x}\pm i\cdot\sigma_{y}$ are Pauli
matrices, $b$ and $b^{+}$ are the annihilation and creation operators
respectively ($[b,b^{+}]=1$). Following the approach of the article \cite%
{17r}, apply the Bogoliubov transformation to the operators $b$ and $b^{+}$ (%
$b(t)=cos(t)\cdot b+sin(t)\cdot b^{+},$ $b^{+}(t)=cos(t)\cdot
b^{+}-sin(t)\cdot b$ ), where $t\in\lbrack0,2\pi]$ is a parameter. We will
use the coherent state representation (Fock-Bargman representation): $b=%
\frac {d}{dz},$ $b^{+}=z$. The eignevalue problem for the operator (\ref{27q}%
) takes the form : 
\begin{equation}
\left[ \frac{\omega_{0}}{2}\left( 
\begin{array}{cc}
0 & 1 \\ 
1 & 0%
\end{array}
\right) +\left( 
\begin{array}{cc}
\overset{\wedge}{c} & 0 \\ 
0 & \overset{\wedge}{c}%
\end{array}
\right) +\left( 
\begin{array}{cc}
\overset{\wedge}{a} & 0 \\ 
0 & -\overset{\wedge}{a}%
\end{array}
\right) \right] \left( 
\begin{array}{c}
\psi_{1} \\ 
\psi_{2}%
\end{array}
\right) =E\left( 
\begin{array}{c}
\psi_{1} \\ 
\psi_{2}%
\end{array}
\right) ,  \label{28q}
\end{equation}
where $\overset{\wedge}{a}=g\left( b\left( t\right) ^{2}+b^{+}\left(
t\right) ^{2}\right) =2g\left( \frac{d^{2}}{dz^{2}}+z^{2}\right) $, $\overset%
{\wedge}{c}=\omega\cdot b^{+}\left( t\right) b\left( t\right) =$ $\frac{%
\omega}{2}\cdot\left[ -\sin\left( 2t\right) \frac{d^{2}}{dz^{2}}+2\cos\left(
2t\right) z\frac{d}{dz}+\sin\left( 2t\right) z^{2}+\cos\left( 2t\right) -1%
\right] $. Removing the function $\psi_{1}\left( z\right) $ from the system (%
\ref{28q}) we come to the fourth-order differential equation for the
function $\psi_{2}\left( z\right) $: 
\begin{equation}
\mathbf{L}\left( t\right) \psi_{2}\left( z\right) \equiv\left[ \frac {1}{%
\omega}\cdot\left( \overset{\wedge}{a}+\overset{\wedge}{c}\right) \left( 
\overset{\wedge}{a}-\overset{\wedge}{c}\right) +\frac{2E}{\omega^{2}}\cdot%
\overset{\wedge}{c}+\frac{\omega_{0}^{2}}{4\omega^{2}}-\frac{E^{2}}{%
\omega^{2}}\right] \psi_{2}\left( z\right) =0  \label{29q}
\end{equation}

Let us try to reduce eq. (\ref{29q}) to the algebraic problem using the
finite-dimensional presentation of the operators $J_{3}^{+},$ $J_{3}^{-}$ (%
\ref{16q}). For this purpose, we shall present the operator $\mathbf{L}%
\left( t\right) $ (\ref{29q}) as polynomial on the operators $J_{3}^{+},$ $%
J_{3}^{-}$ . To this end, we use a gauge transformation $\phi\left( z\right)
^{-1}\mathbf{L}\left( t\right) \phi\left( z\right) =P\left(
J_{3}^{+},J_{3}^{-}\right) |_{x=\mu\left( z\right) }$, where $P\left(
x_{1},x_{2}\right) $ is a polynomial of the second degree in non-commutative
variables $x_{1}$ and $x_{2}$. We restrict ourselves by a special case of $%
\phi\left( z\right) $ and $\mu\left( z\right) $: $\phi\left( z\right)
=\phi_{1}\left( z\right) =\exp\left( \eta\cdot z^{2}\right) $ or $\phi\left(
z\right) =\phi_{2}\left( z\right) =z\cdot\exp\left( \eta\cdot z^{2}\right) $%
, $\mu\left( z\right) =\xi\cdot z^{2}$. Thus, a search of a polynomial
representation for $\mathbf{L}\left( t\right) $ is reduced to a search of
coefficients of polynomial $P\left( x_{1},x_{2}\right) $ and constants $\eta$%
, $\xi$, $t$. The obtained solutions read:

Type I, $s=\frac{1}{2}$, $\alpha=-\frac{1}{4}-\frac{N}{2}$: 
\begin{equation}
\mathbf{L}\left( t_{0}\right) e^{\eta\cdot z^{2}}=\frac{e^{\eta\cdot z^{2}}}{%
3}\left( 2\left( J_{3}^{-}\right) ^{2}+\left[ J_{3}^{+},J_{3}^{-}\right]
-7J_{3}^{+}+4J_{3}^{-}+C_{1}\right) |_{x=-\xi z^{2}}  \label{30q}
\end{equation}

Type II, $s=\frac{3}{2}$, $\alpha=\frac{5}{4}-\frac{N}{2}$: 
\begin{equation}
\mathbf{L}\left( -t_{0}\right) e^{-\eta\cdot z^{2}}=\frac{ze^{-\eta\cdot
z^{2}}}{3}\left( 2\left( J_{3}^{-}\right) ^{2}+\left[ J_{3}^{+},J_{3}^{-}%
\right] -7J_{3}^{+}-8J_{3}^{-}+C_{2}\right) |_{x=\xi z^{2}}  \label{31q}
\end{equation}
where $t_{0}=\frac{1}{4}\arctan\left( \frac{10\sqrt{2}}{23}\right) $, $\frac{%
E}{\omega}=\frac{N+1}{\sqrt{3}}-\frac{1}{2}$, $\eta=\frac{\sqrt{2}}{8}$, $%
C_{1}=3\frac{\omega_{0}^{2}}{4\omega^{2}}-N^{2}-\frac{N}{4}+\frac{1}{8}$, $%
C_{2}=3\frac{\omega_{0}^{2}}{4\omega^{2}}-N^{2}+\frac{13N}{4}+\frac{49}{8}$, 
$\frac{g}{\omega}=\frac{1}{2\sqrt{6}}$, $\xi=\frac{3\sqrt{2}}{8}$. Two types
of the solutions (\ref{30q},\ref{31q}) have different symmetries with
respect to the parity operator $\Pi=\exp\left( i\pi b^{+}b\right) $ \cite%
{17r} ($\left[ \Pi,H\right] =0$).

Further, we will consider the solution (\ref{30q}) (type--I) in more
details, the second solution (\ref{31q}) can be considered in a similar way.
Taking into account the equality (\ref{30q}), the problem (\ref{29q}) can be
reduced to a search of vectors $\phi _{k}$ satisfying the matrix equation $%
\left( 2\left( j_{3}^{-}\right) ^{2}+\left[ j_{3}^{+},j_{3}^{-}\right]
-7j_{3}^{+}+4j_{3}^{-}+C_{1}\right) \phi _{k}=0$, where $j_{3}^{+}$, $%
j_{3}^{-}$ are finite-dimensional representations of the operators $%
J_{3}^{+},J_{3}^{-}$ (\ref{16q}) in the subspace $\mathcal{R}_{N}^{3}$. For
each of $\phi _{k}$, taking into account (\ref{30q}), we have $\mathbf{L}%
\left( t_{0}\right) e^{\eta \cdot z^{2}}\phi _{k}=0$ that is equivalent to (%
\ref{29q}) under the condition $\psi _{2}\equiv \psi _{2}^{k}=e^{\eta \cdot
z^{2}}\phi _{k}$. The functions $\psi _{2}^{k}$ obtained are entire ones and
belong to the Fock space: $\left\vert \psi _{2}^{k}\left( z\right)
\right\vert ^{2}=$ $O\left( \left\vert z\right\vert ^{p}\exp \left( -\frac{%
\sqrt{2}}{4}\left( z^{2}+\overline{z}^{2}\right) \right) \right) $ (type I), 
$\left\vert \psi _{2}^{k}\left( z\right) \right\vert ^{2}=$ $O\left(
\left\vert z\right\vert ^{p}\exp \left( \frac{\sqrt{2}}{4}\left( z^{2}+%
\overline{z}^{2}\right) \right) \right) $ (type II). We list the
corresponding functions explicitly for the special case $N=2$. 
\begin{align}
\text{type I}& \text{: }  \label{32q} \\
\psi _{2}& =e^{\eta \cdot z^{2}}\left( 
\begin{array}{c}
\begin{array}{c}
_{1}F_{1}\left[ \QATOP{-5/4}{1/2};-\xi z^{2}\right] 
\end{array}%
\cdot \left( \frac{57}{20}+\frac{7\sqrt{42}}{15}\right) + \\ 
\begin{array}{c}
_{1}F_{1}\left[ \QATOP{-1/4}{1/2};-\xi z^{2}\right] 
\end{array}%
\cdot \left( \frac{5}{3}+\frac{\sqrt{42}}{30}\right) +%
\begin{array}{c}
_{1}F_{1}\left[ \QATOP{3/4}{1/2};-\xi z^{2}\right] 
\end{array}%
\end{array}%
\right) 
\end{align}%
\begin{equation}
\text{ }\frac{\omega _{0}}{2\omega }=\sqrt{\frac{11}{12}+\frac{\sqrt{42}}{3}}
\label{33q}
\end{equation}%
\begin{align}
\text{type II}& \text{: }  \label{34q} \\
\psi _{2}& =ze^{-\eta \cdot z^{2}}\left( 
\begin{array}{c}
\begin{array}{c}
_{1}F_{1}\left[ \QATOP{1/4}{31/2};\xi z^{2}\right] 
\end{array}%
\cdot \left( \frac{5}{4}+\frac{\sqrt{10}}{7}\right) + \\ 
\begin{array}{c}
_{1}F_{1}\left[ \QATOP{5/4}{3/2};\xi z^{2}\right] 
\end{array}%
\cdot \left( \frac{31}{21}+\frac{\sqrt{10}}{42}\right) +%
\begin{array}{c}
_{1}F_{1}\left[ \QATOP{9/4}{3/2};\xi z^{2}\right] 
\end{array}%
\end{array}%
\right) 
\end{align}%
\begin{equation}
\text{ }\frac{\omega _{0}}{2\omega }=\sqrt{\frac{\sqrt{10}}{3}-\frac{5}{12}}
\label{35q}
\end{equation}%
where $\xi =\frac{3\sqrt{2}}{8}$, $\eta =\frac{\sqrt{2}}{8}$, $\frac{E}{%
\omega }=\sqrt{3}-\frac{1}{2}$. The second component $\psi _{1}$ of the
spectral problem (\ref{28q}) can be obtained from the equation $\psi _{1}=%
\frac{2}{\omega _{0}}\cdot \left( E+\overset{\wedge }{a}-\overset{\wedge }{c}%
\right) \psi _{2}$. In table 1 the values of relative frequencies $\frac{%
2\omega }{\omega _{0}}$ for both types I and II are given for different
dimensions of the subspace $\mathcal{R}_{N}^{3}$. The obtained solutions (%
\ref{30q}, \ref{31q} and table 1) concern a non-resonant case $\frac{2\omega 
}{\omega _{0}}\neq 1$ and are related to the known isolated solutions
(Juddian solution), found earlier in \cite{17r}. It is worth stressing that
in \cite{17r} the eigenfunctions were constructed on the basis of elementary
function, whereas the solutions (\ref{30q}, \ref{31q}) are constructed on
the basis of functions $%
\begin{array}{c}
_{1}F_{1}\left[ \QATOP{\alpha +n}{s};x\right] 
\end{array}%
.$ For the values of the parameters $s=1/2$, $\alpha =-1/4-N/2$ (type I), $%
s=3/2$, $\alpha =5/4-N/2$ (type II) the hypergeometric function does not
degenerate into polynomial ($-\alpha \notin \mathbb{N}\mathtt{)}$.

Table1. Values of parameters Hamiltonian (\ref{27q},\ref{28q}), for
different dim$\left( \mathcal{R}_{N}^{3}\right) $, $\frac{g}{\omega}=\frac{1%
}{2\sqrt{6}}$, $\frac{E}{\omega}=\frac{N+1}{\sqrt{3}}-\frac{1}{2}$.

\begin{tabular}{|c|c|c|c|c|c|c|}
\hline
\multicolumn{2}{|c|}{dim$\left( \mathcal{R}_{N}^{3}\right) $} & 3 & 5 & 6 & 7
& 8 \\ \hline
I & $\frac{2\omega}{\omega_{0}}$ & 0.44315 & 1.68889 & 3.03496 & $%
\begin{array}{c}
\text{2.72766} \\ 
\text{3.60267}%
\end{array}
$ & $%
\begin{array}{c}
\text{2.10305} \\ 
\text{3.74421} \\ 
\text{3.90266}%
\end{array}
$ \\ \hline
II & $\frac{2\omega}{\omega_{0}}$ & 0.79838 & 0.79838 & $%
\begin{array}{c}
\text{2.23006} \\ 
\text{2.75234}%
\end{array}
$ & 3.43545 & $%
\begin{array}{c}
\text{2.66128} \\ 
\text{4.08801}%
\end{array}
$ \\ \hline
\multicolumn{2}{|c|}{$\frac{E}{\omega}$} & 1.23205 & 2.38675 & 2.96410 & 
3.54145 & 4.11880 \\ \hline
\end{tabular}

\section{Conclusion}

To the best of my knowledge, the only previously known example of QES
related to special functions was found in \cite{dz}.where these function
appeared "by chance" for a particular problem connected with quartic Bose
Hamiltonians. Now, we developed a systematic approach of QES-extension that
enables us to generate new QES operators based on special functions. This
extends considerably the family of QES systems and can find physical
applications, one of which (two-photon Rabi Hamilatonian) was discussed in
the present article. The main features of our approach include 1) the
construction of the affine annihilator $\mathcal{K}\left( \mathcal{M}%
_{2}\right) $ \cite{15r}; 2) multiplication at the power function $x^{n}$ .
One can think that the approach suggested in the given work, will give rise
to further essential expansion of classes of quasi exactly solvable models.

\section{Acknowledgments}

The author thanks O. B Zaslavskii for useful discussions.

\section{Appendix A.}

The constructed operators $J_{k}^{+},$ $J_{k}^{-}$ $(k=1\ldots6)$, satisfy
to the following commutation relations: 
\begin{equation}
\left[ J_{k}^{-},J_{k}^{+}\right] =S_{k}  \label{A1}
\end{equation}%
\begin{equation}
\left[ J_{k}^{+},S_{k}\right] =c_{1}^{+}\cdot\left( J_{k}^{-}\right)
^{2}+c_{3}^{+}\cdot J_{k}^{+}J_{k}^{-}+c_{4}^{+}\cdot S_{k}+c_{5}^{+}\cdot
J_{k}^{+}+c_{6}^{+}\cdot J_{k}^{-}+c_{7}^{+}  \label{A2}
\end{equation}%
\begin{equation}
\left[ J_{k}^{-},S_{k}\right] =c_{1}^{-}\cdot\left( J_{k}^{-}\right)
^{2}+c_{2}^{-}\cdot\left( J_{k}^{+}\right) ^{2}+c_{5}^{-}\cdot
J_{k}^{+}+c_{6}^{-}\cdot J_{k}^{-}+c_{7}^{-}  \label{A3}
\end{equation}
where $k=1\ldots6$, $c_{i}^{\pm}$ $(i=1,\ldots,7)$ are constants, theirs
value are given in the table 2. The operators $S_{k}$ (\ref{A1}) have the
general structure $S_{k}=\beta_{3}\left( x\right) \frac{d^{3}}{dx^{3}}+\beta
_{2}\left( x\right) \frac{d^{2}}{dx^{2}}+\beta_{1}\left( x\right) \frac {d}{%
dx}+\beta_{0}\left( x\right) $, $\beta_{m}\left( x\right) \in\mathcal{P}_{n}$
$(m=0,$ $1,$ $2,$ $3)$. However, we do not list them here explicitly since
the corresponding expressions are rather cumbersome.

Table2. Values of constants $c_{i}^{\pm}$ $(i=1,\ldots,7)$ included in the
commutation relations (\ref{A1}-\ref{A3}), $C_{N}=2+2\alpha+s$, $\alpha
_{N}=2+N+2\alpha$, $\beta_{N}=-4\cdot\left( 1+\nu^{2}+2N+\nu\right) $, $%
\gamma_{N}=\left( \alpha-N\right) \cdot\left( s+1\right) \cdot C_{N}$, $%
\delta_{N}=-8\cdot\left( 2\alpha-N\right) \cdot\alpha_{N}$, $%
S_{N}=s\cdot\alpha\cdot\left( s-N-2\right) $, $A_{N}=s+N+2\alpha$, $%
B_{N}=\left( 2N-s\right) \cdot\left( s-2-2N\right) $, $D_{N}=\left(
s+1\right) \cdot\left( s-2-2N\right) $, $G_{N}=\left( \alpha-N\right)
\cdot\left( s+1\right) $.

\begin{tabular}{|c|c|c|c|c|c|c|}
\hline
& $k=1$ & $k=2$ & $k=3$ & $k=4$ & $k=5$ & $k=6$ \\ \hline
$c_{1}^{+}$ & $0$ & $0$ & $0$ & $0$ & $\ 0$ & $-6$ \\ \hline
$c_{1}^{-}$ & $2$ & $2$ & $2$ & $0$ & $2$ & $0$ \\ \hline
$c_{2}^{-}$ & $0$ & $0$ & $0$ & $6$ & $0$ & $0$ \\ \hline
$c_{3}^{+}$ & $-4$ & $-4$ & $-4$ & $0$ & $-4$ & $0$ \\ \hline
$c_{4}^{+}$ & $-2$ & $-2$ & $-2$ & $0$ & $-2$ & $0$ \\ \hline
$c_{5}^{+}$ & $2$ & $C_{N}$ & $A_{N}$ & $-2$ & $0$ & $0$ \\ \hline
$c_{5}^{-}$ & $0$ & $1$ & $1$ & $0$ & $4$ & $4$ \\ \hline
$c_{6}^{+}$ & $B_{N}$ & $B_{N}$ & $s-N$ & $0$ & $1-4N^{2}$ & $12+32\alpha $
\\ \hline
$c_{6}^{-}$ & $-2$ & $-C_{N}$ & $-A_{N}$ & $2$ & $0$ & $0$ \\ \hline
$c_{7}^{+}$ & $D_{N}$ & $\gamma_{N}$ & $S_{N}$ & $0$ & $0$ & $\delta_{N}$ \\ 
\hline
$c_{7}^{-}$ & $0$ & $G_{N}$ & $s\cdot\alpha$ & $0$ & $\beta_{N}$ & $0$ \\ 
\hline
\end{tabular}


\begin{thebibliography}{99}
\bibitem{1r} O. B. Zaslavskii, and V. V. Ulyanov, Sov. Phys. JETP \textbf{60,%
} 991 (1984).

\bibitem{2r} O. B. Zaslavskii, and V. V. Ulyanov, Theor. Math. Phys. \textbf{%
71,} 520 (1987).

\bibitem{3r} M. A. Shifman, Int. J. Mod. Phys.\textit{\ A} \textbf{4,} 3305
(1989).

\bibitem{4r} O. B. Zaslavskii, \textit{Sov. Phys. J.} \textbf{33,} 12 (1990).

\bibitem{5r} M. A. Shifman, and A. V. Turbiner, Commun. Math. Phys. \textbf{%
126,} 347 (1989).

\bibitem{6r} A. V. Turbiner, Commun. Math. Phys. \textbf{118,} 467 (1988).

\bibitem{7r} A. Ushveridze, \textit{Quasi-Exactly Solvable Models in Quantum
Mechanics} (Bristol: Intitute of Physics Publishing, 1994).

\bibitem{8r} V. V. Ulyanov, and O. B. Zaslavskii, Phys. Rep. \textbf{216,}
179 (1992).

\bibitem{9r} D. Gomez-Ullate, N. Kamran, and R. Milson, \textit{Preprint}
arXiv nlin.SI/0601053, (2006).

\bibitem{10r} D. Gomez-Ullate, N. Kamran and R. Milson, \textit{Preprint}
arXiv nlin.SI/0610065, (2006).

\bibitem{11r} G. Post, and A. Turbiner, \textit{Preprint} arXiv
funct-an.SI/9307001, (1993).

\bibitem{12r} Y. Brihaye, \textit{Preprint} arXiv math-ph/0401005, (2004).

\bibitem{13r} Y. Brihaye, J. Ndimubandi, and B. P. Mandal \textit{Preprint}
arXiv math-ph/0601004, (2006).

\bibitem{14r} R. Z. Zhdanov, J. Math. Phys. \textbf{37,} 3198 (1996).

\bibitem{15r} N. Kamran, R. Milson, and P. J. Olver, Adv.in Math. \textbf{%
156,} 286 (2000).

\bibitem{16r} M. Abramowitz, and I. A. Stegun (eds) \textit{Handbook of
Mathematical Functions} (New York:Dover, 1970) or
http://functions.wolfram.com/HypergeometricFunctions/ , http://functions.
wolfram.com/BesselAiryStruveFunctions/

\bibitem{17r} C. Emary and R. F. Bishop, J. Phys. A: Math. Gen. \textbf{35,}
8231 (2002).

\bibitem{18r} I. I. Rabi, Phys. Rev. \textbf{51,} 652 (1937).

\bibitem{dz} S. N. Dolya, and O. B. Zaslavskii, J. of Phys. A: Math. Gen. 
\textbf{34,} 5995 (2001).
\end{thebibliography}
\end{document}